\documentclass[aps,prd,preprint,superscriptaddress,showpacs,floatfix,preprintnumbers,nofootinbib,a4paper]{revtex4-1}
\usepackage{graphicx}
\usepackage{color}
\usepackage{graphics}
\usepackage{subfigure}
\usepackage{amsmath, amssymb, epsfig}
\usepackage{textcomp}
\newcommand{\mathsym}[1]{{}}

\newcommand{\ba}{\begin{array}}
\newcommand{\ea}{\end{array}}
\newcommand{\be}{\begin{equation}}
\newcommand{\ee}{\end{equation}}
\newcommand{\beqa}{\begin{eqnarray}}
\newcommand{\eeqa}{\end{eqnarray}}
\def\321{$SU(3)\times SU(2)\times U(1)$}

\def\vev#1{\left\langle #1\right\rangle}%

\begin{document}
\vspace*{1cm}
\title{Horizontal symmetries of leptons with a massless neutrino}
\bigskip
\author{Anjan S. Joshipura}
\email{anjan@prl.res.in}
\affiliation{Physical Research Laboratory, Navarangpura, Ahmedabad 380 009, India.}
\author{Ketan M. Patel}
\email{ketan@theory.tifr.res.in}
\affiliation{Department of Theoretical Physics, Tata Institute of Fundamental Research, Mumbai
400 005, India.\vspace*{1cm}}

\preprint{TIFR/TH/13-19}

%--------------------------------------------------------------
\begin{abstract}
\vspace*{0.2cm}
Residual symmetry $G_\nu$ of neutrino mass matrix with a massless neutrino and embedding of $G_\nu$
and the residual symmetry $G_l$ of the charged lepton mass matrix into  finite discrete groups $G$
is discussed. Massless neutrino results if $G_\nu$ and hence $G$ are subgroups of $U(3)$ rather than
of  $SU(3)$. Structure of the resulting leptonic
mixing matrix $U_{PMNS}$ is  discussed in three specific examples based on groups (a) $\Sigma(3N^3)$, (b) $\Sigma(2N^2)$
and (c) $S_4(2) \equiv A_4\rtimes Z_4$. $\Sigma(3N^3)$ groups are able to reproduce either the second or the third
column of $U_{PMNS}$ correctly. $\Sigma(2N^2)$ groups lead to prediction $\theta_{13}=0$, $\theta_{23}=\frac{\pi}{4}$
for the reactor and atmospheric mixing angles respectively if neutrino mass hierarchy is inverted. Solar angle remains
undetermined in this case. This also gets determined when $G=S_4(2)$ which can give bi-maximal mixing for inverted hierarchy. Examples (b) and (c) provide a good zeroth order approximation to realistic leptonic mixing with a massless neutrino. We also present an example of the specific model based on $S_4(2)$ symmetry in which a massless neutrino and viable leptonic mixing angles are obtained.
\end{abstract}
%----------------------------------------------------------

\maketitle

\section{Introduction}
The observed leptonic mixing angles \cite{GonzalezGarcia:2012sz,*Fogli:2012ua,*Tortola:2012te} are known to be close to
special values. The atmospheric mixing angle is close to maximal with $\sin^2\theta_{23}\approx0.44 \sim \frac{1}{2}$,
the solar angle $\theta_{12}$ and the reactor angle $\theta_{13}$ satisfy $\sin^2\theta_{12}\approx 0.31\sim
\frac{1}{3}$ and $\sin^2\theta_{13}\approx 0.023\sim 0$. It is natural to look for  group theoretical explanations
for such special values as has been extensively done, see
\cite{Altarelli:2010gt,*Altarelli:2012ss,*King:2013eh,Ishimori:2010au} for reviews. In this approach, it is assumed that
underlying theory of leptonic flavor possesses some discrete symmetry $G$. The group $G$ breaks to smaller non-commuting 
subgroups $G_\nu$ and $G_l$ which correspond to unbroken symmetries respectively of the neutrino and the charged lepton mass
matrices $M_\nu$ and $M_l$, more precisely of $M_lM_l^\dagger$. While possible choices of $G$ are a priori unknown and
numerous, one can relate $G_\nu$ and $G_l$ \cite{Lam:2007qc,Lam:2008rs,*Lam:2008sh,Grimus:2009pg} to the known structure
of the mixing matrix. Thus it becomes more profitable to start with possible choices of $G_\nu$ and
$G_l$ dictated from physical considerations and search for groups which contain them as subgroups.
In this way, Lam \cite{Lam:2008rs,*Lam:2008sh} argued that minimal group which combines symmetries
of $M_\nu$ with tri-bimaximal (TBM)
mixing \cite{Harrison:2002er,*Xing:2002sw} structure and a diagonal $M_lM_l^\dagger$ is $S_4$.

In all these analysis, basic but implicit assumption is that neutrinos are Majorana particles and all three of them 
are massive\footnote{A recent analysis in \cite{Hernandez:2013vya} addresses the problem of relating mixing to
symmetries in case of completely or partially degenerate neutrino masses.}. The present neutrino data are however quite
consistent with one of the neutrinos being exactly massless both in case of the normal and inverted hierarchy for
neutrino masses. The underlying symmetry $G_\nu$ and hence possible choice of $G$ become quite different in this case. 
In this note, we discuss possible symmetry groups $G_\nu$ and embedding of $G_\nu$ and $G_l$ into some bigger group $G$
assuming that one of the three neutrinos is  massless.

Let us quickly summarize the steps
\cite{Lam:2007qc,Lam:2008rs,*Lam:2008sh,Toorop:2011jn,Hernandez:2012ra,*Hernandez:2012sk,Holthausen:2012wt} used in
relating mixing angles to symmetry groups $G$. Let $U_\nu$ and $U_l$ respectively diagonalize
$M_\nu$ and 
$M_lM_l^\dagger$:
\beqa \label{unu}
U_\nu^T M_\nu U_\nu&=&{\rm Diag.}(m_{\nu_1},~m_{\nu_2},~m_{\nu_3})~, \nonumber \\
U_l^\dagger M_l M_l^\dagger U_l &=& {\rm Diag.}(m_e^2,~m_\mu^2,~m_\tau^2)~.\eeqa
Assume that $M_\nu$ $(M_lM_l^\dagger)$ is invariant under some set of discrete symmetries $S_i$ $(T_l)$:
\be \label{snu}
S_i^T M_\nu S_i=M_\nu~~~{\rm and}~~~T_l^\dagger M_lM_l^\dagger T_l=M_lM_l^\dagger~.\ee
It is assumed that elements within $S_i$ and $T_l$  commute among themselves and hence can be simultaneously
diagonalized by unitary matrices $V_\nu$ and $V_l$ respectively:
\be \label{vnu}
V_\nu^\dagger S_i V_\nu=s_i~~~{\rm and}~~~ V_l^\dagger T_l V_l=t_l~,\ee
where $s_i$ and $t_l$ correspond to diagonal matrices. Eqs. (\ref{unu}, \ref{snu}, \ref{vnu}) can be
used to show that
\cite{Lam:2007qc,Lam:2008rs,*Lam:2008sh,Toorop:2011jn,Hernandez:2012ra,*Hernandez:2012sk,Holthausen:2012wt}
\be \label{relation-UV}
U_\nu=V_\nu P_\nu~~~{\rm and}~~~U_l=V_l P_l~, \ee
where $P_{l, \nu}$ are diagonal phase matrices. Therefore, one can write:
\be \label{relation}
U \equiv U_{PMNS} = U_l^\dagger U_\nu= P_l^*V_l^\dagger V_\nu P_\nu~.\ee
Note that $S_i$ and $T_l$ denote the 3-dimensional representations of elements of some symmetry
group $G$ in this
approach. The structure of these symmetries and the matrices $V_{l,\nu}$ diagonalizing them is thus determined by group
theory and Eq. (\ref{relation}) provides a direct link between leptonic mixing and group theory.

In a bottom up approach, one first determines groups of $S_i$ and $T_l$ and then uses them to find suitable group
$G$. A complete set of $S_i$ and $T_l$ may depend on underlying dynamics. However one can  define a minimal set
which can always be taken as symmetries of mass matrices. A field corresponding to a massive Majorana neutrino is
arbitrary up to a change of sign in the mass basis. If all three neutrinos are massive then the corresponding diagonal
mass matrix is trivially invariant under
\be \label{snu0}
s_1={\rm Diag.}(1,~-1,~-1)~,~~s_2={\rm Diag.}(-1,~1,~-1)~~{\rm and}~~s_3=s_1 s_2~, \ee
where Det$(s_{i})$ is chosen +1. Any two of these define a $Z_2\times Z_2$ symmetry. One can go to arbitrary basis and
define corresponding $G_\nu =Z_2\times Z_2$ symmetry transformation $S_i=V_\nu s_i V_\nu^\dagger$ as a symmetry of the
general neutrino mass matrix. The fields corresponding to the charged lepton mass eigenstates are invariant under three
independent $U(1)$ symmetries and Eq. (\ref{snu0}) gets replaced by 
\be \label{sl0} 
t_l={\rm Diag.}(e^{i\phi_e},~e^{i\phi_\mu},~e^{i\phi_\tau})~~.\ee
Assuming that $t_l$ is an element of some discrete group, the phases $\phi_{e,\mu,\tau}$ would be restricted to some 
discrete values and the most general $G_l$ would be a discrete sub-group of $U(1)^3$.  Conversely, one can start with a
group $G$, identify its $Z_2\times Z_2$ subgroup corresponding to $G_\nu$ and appropriate $G_l$ and
use them to predict
the observed mixing. In this way, $G=A_4$, $S_4$, $A_5$, $PSL(2,Z_7)$, $\Delta(96)$, $\Delta(384)$
\cite{Altarelli:2010gt,*Altarelli:2012ss,*King:2013eh,Ishimori:2010au,deAdelhartToorop:2011re},
$\Delta(150)$, $\Delta(600)$ \cite{Lam:2013ng} and general $\Delta(6n^2)$ \cite{King:2013vna} are studied for
their predictions of the mixing angles. A complete scan over large number of groups is performed 
\cite{Parattu:2010cy,Holthausen:2012wt} and it is found that only three of about a million groups analyzed in 
\cite{Holthausen:2012wt} can predict all the mixing angles within 3$\sigma$. However there exist many choices which
lead to very good zeroth order approximation. In particular, groups leading to democratic, bi-maximal (BM) or a
TBM mixing matrix are identified. A summary of various cases is given in
\cite{deAdelhartToorop:2011re}. Alternative approach is also proposed in which one  relates mixing matrix elements
directly to group theoretical parameters using various von-Dyck groups
\cite{Hernandez:2012ra,*Hernandez:2012sk,Hu:2012ei,*Lam:2013xs,*Grimus:2013rw}.

\section{Residual symmetry with a massless neutrino}
Let us now discuss the situation when one of the neutrinos is massless. It follows from Eq. (\ref{snu}) that  
Det$(M_\nu)=0$ if Det$(S_i)\neq\pm 1$. Thus if neutrino mass matrix is invariant under an element of
a subgroup of 
$U(3)$ which is not in   $SU(3)$ then such invariance automatically implies the presence of at least one massless state.
The underlying group $G$ containing $S_i$ necessarily belongs to $U(3)$ and one must look for  groups different from 
the ones used in the existing studies \cite{Altarelli:2010gt,*Altarelli:2012ss,*King:2013eh}.

Residual symmetry of neutrino has to be very specific if one further requires that only one of the neutrinos is
massless. While corresponding mass eigenstate appears in the definition of flavor eigenstates and hence in Lagrangian,
the neutrino mass term is trivially invariant under a $U(1)$  symmetry which corresponds to an arbitrary change of phase
of the massless field. In addition, one can independently change the signs of other two massive states which correspond
to a $Z_2\times Z_2$ symmetry. Thus the full residual symmetry of the neutrino mass matrix is now $Z_2\times Z_2\times
U(1)$  instead of $Z_2\times Z_2$. We shall restrict ourselves to a discrete subgroup $Z_N$ (with $N\geq3$) of $U(1)$
and take the residual symmetry as $Z_2\times Z_2\times Z_N$ with the following definition for $Z_N$ in an arbitrary
basis
\be \label{snugen}
S=V_\nu~{\rm Diag.}\left(e^{2\pi i \frac{k}{N}},~1,~1\right) ~V_\nu^\dagger~.\ee
with $k=1,2,..,N-1$ and $e^{2\pi i \frac{k}{N}}\neq-1$. This describes normal hierarchy while the inverted hierarchy can
be obtained by replacement $V_\nu \rightarrow V_\nu P_{13}$ in the above $S$, where $P_{13}$ is a permutation matrix in
1-3 plane.

We start with an example which brings out clear differences between situation with a massless state compared to 
all three neutrinos being massive. Consider the following $Z_2$ and $Z_3$ as residual symmetries respectively for the
neutrino and the charged lepton mass matrices:
\be \label{snut}
s_1=\left(
\ba{ccc}
1&0&0\\
0&-1&0\\
0&0&-1\\ \ea\right)~~~{\rm and}~~~
T=\left( 
\ba{ccc}
0&1&0\\
0&0&1\\
1&0&0\\ \ea
\right)~.\ee
It is known \cite{Lam:2007qc,Grimus:2009pg} that $s_1$ and $T$ together generate the $A_4$ group with presentations
$S=s_1$ and $T$ satisfying $S^2=T^3=(ST)^3=1$. The leptonic mixing matrix $U$ in this case
can be worked out using
Eq. (\ref{relation}) and is given by the following apart from phase matrices.
\be \label{a4mixing}
U\equiv U_\omega^\dagger U_{23}(\theta) ~,\ee
where 
\be \label{uomega}
U_{\omega}=\frac{1}{\sqrt{3}} \left(
\ba{ccc}
1&1&1\\
1&\omega^2 &\omega\\
1&\omega &\omega^2\\
\ea \right)\ee
with $\omega^3=1$ diagonalizes $T$ and $U_{23}(\theta)$ is a unitary rotation in 2-3 plane arising due to degeneracy of
two eigenvalues in $s_1$. The mixing matrix is democratic if $\theta=0$. For $\theta=\pi/4$, absolute values of columns
of $U$ coincide with columns of $|U_{TBM}|$ describing the TBM mixing. Whether such $U$ can describe a good zeroth order
approximation to the observed mixing further depends on the neutrino mass hierarchy which is not predicted by group
theory and depends on the model parameters. There exist variety of models based on $A_4$ giving correct hierarchy and
TBM \cite{Altarelli:2010gt,*Altarelli:2012ss,*King:2013eh}. Note that $\theta$ is a free parameter if only $A_4$ is
used. It can be fixed at a value $\pi/4$ by adding another $Z_2$ symmetry corresponding to
$\mu$-$\tau$ interchange on the
neutrino mass matrix. The group containing $s_1$, $T$ and $\mu$-$\tau$ interchange symmetry is $S_4$
\cite{Lam:2008rs,*Lam:2008sh}.

Let us now demand one of the states to be massless and replace $s_1$ of Eq. (\ref{snut}) with an analogous symmetry
\be\label{sigma81}
s_1(\omega) =\left(
\ba{ccc}
\omega&0&0\\
0&1&0\\
0&0&1\\ \ea\right)~,\ee
The multiple products of $s_1(\omega)$ as given in Eq. (\ref{sigma81}) and $T$ as given in Eq. (\ref{snut}) together
form a finite group $\Sigma(81)$. This is seen as follows. Define, $a''\equiv s_1(\omega)$, $b\equiv T$, $a'\equiv
T^{-1}a'' T$ and $a\equiv T^{-1}a' T$. The matrices $a$, $a'$, $a''$, $b$ define \cite{Ishimori:2010au} generators of
the group $\Sigma(81)$ whose elements are labeled as $g=b^k a^l a'^m a''^{n}$ with $k,l,m,n=0,1,2$. $\Sigma(81)$ is
known to be a subgroup of $U(3)$ and not of $SU(3)$
\cite{Ishimori:2010au,Grimus:2011fk,Ludl:2010bj}. It has been
used as a flavor symmetry in \cite{Ma:2007ku,*Hagedorn:2008bc,*BenTov:2012xp} entirely for different reasons. While the
group $A_4$ obtained in case with all massive neutrinos gets replaced by $\Sigma(81)$, the leptonic mixing matrix is
still formally given by Eq. (\ref{a4mixing}) obtained in case of $A_4$. But unlike in case of $A_4$, now the neutrino
mass hierarchy is also partly determined by the symmetry due to one vanishing mass. By construction, the massless state
in the mass basis corresponds to an eigenvector $|\psi_0\rangle=(1,~0,~0)^T$ which becomes in the flavor basis 
$U|\psi_0\rangle=\frac{1}{\sqrt{3}}(1,~1,~1)^T$ independent of $\theta$ appearing in Eq. (\ref{a4mixing}). This
state being massless cannot  be associated with heavier of the solar pair and above state must correspond to the first
column of $U$ and not the second column. Thus the TBM pattern cannot be realized in this
simple example.

\section{Examples of mixing patterns with a massless neutrino}
We now follow the strategy as given in the previous section and study several examples of groups accommodating a 
massless neutrino and derive various mixing patterns implied by them. While all finite subgroups of $SU(3)$ are
systematically classified, see \cite{Grimus:2010ak} for a review, not all subgroups of $U(3)$ are known. Ref. 
\cite{Parattu:2010cy} has listed all such subgroups of order less than 100 and this analysis is extended in
\cite{Ludl:2010bj} where all finite subgroups of $U(3)$ of order 512 or less possessing a faithful three dimensional
irreducible representation are listed. In the following, we consider three different class of examples based on groups
$\Sigma(3 N^3)$, $\Sigma(2 N^2)$ and $S_4(2)\equiv A_4\rtimes Z_4$ and study possible mixing patterns implied by them.

\subsection{Mixing pattern with $\Sigma(3N^3)$}
The example of $\Sigma(81)$ studied above admits a straightforward generalization to a  group series $\Sigma(3N^3)$.
Properties of these groups are studied in \cite{Grimus:2011fk,Ishimori:2010au}. These groups are semi-direct product of
product $Z_N\times Z_N'\times Z_N''$ of three cyclic groups  with a $Z_3$ group. Adopting notation of
\cite{Ishimori:2010au}, the $Z_N$ groups are generated by $a$, $a'$, $a''$ satisfying 
$a^N=a'^N=a''^N=1$ and commuting with each other. The $Z_3$ generator corresponding to cyclic
permutation transforms $a$, $a'$ and $a''$  among each other as follows:
\be \label{sigma3n}
b^{-1} a b=a'',~~b^{-1}a''b=a'~~{\rm and}~~b^{-1} a'b=a~.\ee
As a result of the above equation, all $3N^3$ elements of the group can be written as $g(k,l,m,n)= b^k a^l a'^m a''^n$
with $k=0,1,2$ and $l,m,n=0,1,...,N-1$.

A specific 3-dimensional representation for the generators is given by
\be \label{generators}
a=\left(
\ba{ccc}
1&0&0\\
0&1&0\\
0&0&\rho\\
 \ea\right),~
a'=\left(
\ba{ccc}
1&0&0\\
0&\rho&0\\
0&0&1\\
 \ea\right),~
a''=\left(
\ba{ccc}
\rho&0&0\\
0&1&0\\
0&0&1\\
 \ea\right),~
b=\left(
\ba{ccc}
0&1&0\\
0&0&1\\
1&0&0\\
 \ea\right),\ee
with $\rho=e^{2\pi i/N}$. Group elements $g(0,l,m,n)= a^l a'^m a''^n$ are diagonal in this
specific
representation. From these one can choose $g(0,0,0,p)$ as the residual symmetry of the neutrino mass matrix with
$p=1,2,..,N-1$. The
residual symmetry of the charged lepton has to be non-diagonal and only allowed choices are $g(1,l,m,n)$ and
$g(2,l,m,n)$. We choose $g(1,l,m,n)$ but other choice also gives identical mixing pattern. Let's choose    
\be\label{st3ncube}
S=\left( \ba{ccc}
\rho^{p}&0&0\\
0&1&0\\
0&0&1\\
\ea \right)~~~{\rm and}~~~
T=\left( \ba{ccc}
0&\rho^m&0\\
0&0&\rho^l\\
\rho^n&0&0\\
\ea \right) ~.\\ \ee
Now $T$ is diagonalized by a unitary matrix $U_\rho$ such that $U_\rho^\dagger T U_\rho=
{\rm Diag.}(\lambda_1,~\lambda_2,~\lambda_3)$, where 
\be \label{urho}
U_\rho=\frac{1}{\sqrt{3}}\left( \ba{ccc}
1&1&1\\
\lambda_1\rho^{-m}&\lambda_2\rho^{-m}&\lambda_3\rho^{-m}\\
\lambda_1^2\rho^{-l-m}&\lambda_2^2\rho^{-l-m}&\lambda_3^2\rho^{-l-m}\\
\ea \right) ~.\ee
Here, the eigenvalues $\lambda_i~~(i=1,2,3) $ satisfy
\beqa
\lambda_i^3&=&\rho^{m+n+l} ~,\nonumber\\
\lambda_1+\lambda_2+\lambda_3&=&0~,\nonumber\\
\lambda_1^2+\lambda_2^2+\lambda_3^2&=&0~.\eeqa
The solutions to these equations are
\be \label{sols}
\{\lambda_1,~\lambda_2,~\lambda_3\} = \delta\{1,~\omega^2,~\omega\}~~{\rm or}~~\delta\{1,~\omega,~\omega^2\}~.\ee
where $\delta^3=\rho^{n+m+l}$ and $\omega^3=1$. Thus one can rewrite all possible $U_\rho$ as
\be \label{urho-sol}
U_\rho = P_\delta U_\omega~~~~{\rm or}~~~~U_\rho = P_\delta U_{\omega}^*,\ee
where $P_\delta={\rm Diag.}(1,~\delta\rho^{-m},~\delta^2\rho^{-l-m})$ and $U_\omega $ is given in
Eq. (\ref{uomega}).
Given the above, one can now work out the predicted mixing pattern using Eq. (\ref{relation}): 
\be \label{pmns3ncube}
U=U_\rho^\dagger U_{23}(\theta)~.\ee
This is very similar to the structure obtained in case of $\Sigma(81)$ which is a special case of these groups with 
$N=3$. Once again, the state which remains massless is given by $|\psi_0\rangle=(1,~0,~0)^T$ in the mass basis. The 
 composition of the corresponding state in the flavor basis
$|\psi_f\rangle\equiv U|\psi_0\rangle=U_{\omega}^\dagger|\psi_0\rangle$ is independent of the angle and phase in
$U_{23}$. Also, it does not depend on the charges $l$, $m$, $n$ of the residual symmetry of
the charged leptons since
$P_\delta^*|\psi_0\rangle=|\psi_0\rangle$. As a result, one finds $|\psi_f\rangle=\frac{1}{\sqrt{3}}(1,~1,~1)^T$
from Eqs. (\ref{urho}, \ref{urho-sol}).

The choice of $S$ in Eq. (\ref{st3ncube}) leads to normal hierarchy for which  the state $|\psi_f\rangle$ has to be
identified with the first column of $U$. The second and the third columns depend on the choice of angle $\theta$ and one
has two possibilities. (1) $\theta$ is such that  $|U_{e2}|$ is also $\frac{1}{\sqrt{3}}$ so that one gets the TBM
value.  Then by orthogonality one also predicts $|U_{e3}|=\frac{1}{\sqrt{3}}$ which is far from the observed value. 
(2) $\theta$ may be chosen to get $|U_{e3}|=0$.  Then one automatically gets $\sin^2\theta_{12}=\frac{2}{3}$. Both these
choices are possible. In general, Eq. (\ref{pmns3ncube}) leads to
\beqa \label{Sigma3N-results}
\sin\theta_{13}&=& \frac{1}{\sqrt{3}}|s - c \delta^* \rho^l|~,\nonumber \\
\cos\theta_{13}\sin\theta_{12}&=& \frac{1}{\sqrt{3}}|c + s \delta^* \rho^l|~,\nonumber \\
\cos\theta_{13}\sin\theta_{23}&=& \frac{1}{\sqrt{3}}|s -c \omega \delta^* \rho^l|~, \eeqa
where $s=\sin\theta$, $c=\cos\theta$ and we have neglected phase in $U_{23}$ for simplicity. It does not alter the
result. For the second solution in Eq. (\ref{urho-sol}), $\omega^2$ replaces $\omega$ in the last expression in Eq.
(\ref{Sigma3N-results}). Now if we require $\sin\theta_{13}=0$ in above equation then $s=c$ and $\delta=\rho^l$. This
then implies 
$$ \sin^2\theta_{12}=2/3~~~~{\rm  and}~~~~\sin^2\theta_{23}=1/2~. $$
By deviating away from $s=c$ or $\delta=\rho^l$ one either generates large $\theta_{13}$ or small deviation in
prediction for the solar mixing angle. Above equations (or equivalently composition of the massless state) can be 
used to show that irrespective of the choice of $\theta$, $\delta$ and $\rho$ one has 
$\sin^2\theta_{12} \cos^2\theta_{13} + \sin^2\theta_{13}=\frac{2}{3}$. This  substantially differs from the
experimentally required value $\approx \frac{1}{3}$. The choice $s=0$ gives democratic mixing which is also found to
emerge from other groups \cite{deAdelhartToorop:2011re} like $A_4$ in the presence of all three massive neutrinos.

The same argument holds even if one chooses $g(2,l,m,n)$ as $T$ or $S$ is chosen with a phase in either (2,2) or (3,3)
entry instead of Eq. (\ref{st3ncube}). In all these cases, structure of the massless state is independent of the unknown
angle $\theta$ of the mixing matrix and has a trimaximal form with equal mixture of all three flavours in it. Thus
this class of groups can predict either the third or the second column of the PMNS matrix correctly but not both of
them simultaneously. We now turn to another choices which can lead to mixing patterns close to reality.

\subsection{Mixing pattern with $\Sigma(2N^2)$}
The example in this section is based on the $\Sigma(2 N^2)$ groups whose properties are listed
in \cite{Ishimori:2010au}. Unlike many examples considered in literature
\cite{Altarelli:2010gt,*Altarelli:2012ss,*King:2013eh} 
these groups do not admit  a faithful 3-dimensional irreducible representation and we will use reducible $2+1$ dimensional representation to describe leptonic doublets. As we will see, this too leads to a realistic mixing pattern modulo one free parameter. 
 
$\Sigma(2 N^2)$ is  constructed from two $Z_N$ and a $Z_2$ group and it is isomorphic to $Z_N\times Z_N'\rtimes Z_2$.
Two commuting $Z_N$ generators $a$ and $a'$ satisfy
$$ a^N=a'^N=1~.$$
$Z_2$ generator $b$ transforms them into each other:   
\be 
b^2=1~~{\rm and}~~ bab=a'. \ee
Because of the above defined semi-direct product structure all the groups elements can be written as $b^k a^p a'^{q}$ 
with $k=0,1$ and $p,q=0,1,..,N-1$. All the irreducible representations of the group are either 1 or 2 dimensional. 
The two dimensional representations are labeled by two $Z_N$ charges $(p,q)$. One can represent the generators in a
2-dimensional representation $(1,0)$ \cite{Ishimori:2010au} by
\be \label{sgm2n-gen}
a=\left( \ba{cc}
1&0\\ 0&\rho\\ \ea \right),~~
a'=\left( \ba{cc}
\rho&0\\ 0& 1\\ \ea \right)~~{\rm and}~~
b=\left( \ba{cc}
0&1\\ 1& 0\\ \ea \right)~,\ee
where $\rho=e^{2\pi i /N}$. Using the above representation all the elements of $\Sigma(2N^2)$ can be written as
\cite{Ishimori:2010au}
\be \label{sgm2n-elements-2d}
s_{mn}=\left( \ba{cc} \rho^m&0\\ 0&\rho^n\\ \ea \right)~~{\rm and}~~
t_{mn}=\left( \ba{cc} 0 & \rho^m\\ \rho^n& 0\\ \ea \right),\ee
where $m,n=0,1,...,N-1$. The elements $t_{mn}$ in the above equation can be diagonalized as $v_\rho^\dagger t_{mn}
v_\rho={\rm Diag.}(\lambda_1,~\lambda_2)$. Such $v_\rho$ can be written as
\be \label{sgm2n-U}
v_\rho=\frac{1}{\sqrt{2}}\left( \ba{cc}
1&1\\ \lambda_1\rho^{-m}& \lambda_2\rho^{-m}\\ \ea \right),\ee
where $\lambda_1 = -\lambda_2 = \sqrt{\rho^{m+n}}$.

Let's now apply this symmetry to the lepton sector by constructing its 3-dimensional reducible representation from the
above. Let's define 
\be \label{sgm2n}
S_{qp}=\left( \ba{ccc} 1&0&0\\ 0&\rho^q&0\\ 0&0&\rho^p\\ \ea \right)~~{\rm and}~~
T_{mn}=\left( \ba{ccc} 1&0&0\\0 &0&\rho^m\\ 0&\rho^n&0\\ \ea \right)\ee
$S_{0p}$ would be the appropriate   symmetry of the neutrino mass matrix with inverted hierarchy. In order to get
non-trivial mixing pattern, the charged lepton mass matrix should posses a symmetry which is generated by any of
$T_{mn}$ such that
\be \label{symm}
S_{0p}^T M_\nu S_{0p} = M_\nu~~{\rm and}~~T_{mn}^\dagger M_l M_l^\dagger T_{mn} = M_l M_l^\dagger.\ee
The above symmetry ensures $m_{\nu_3}=0$ for $N\geq3$. The $3\times 3$ matrix $V_\rho$ diagonalizing $T_{mn}$ has a
block diagonal form with the lower $2\times 2$ block given by the matrix $v_\rho$ in Eq. (\ref{sgm2n-U}) and
$(V_\rho)_{11}=1$. Once again using Eq. (\ref{relation}), the PMNS matrix can be written as 
\be \label{Umns-inverted}
U = V_\rho^\dagger U_{12}(\theta_\nu) = R_{23}(\pi/4)U_{12}(\theta_\nu)P_\nu, \ee
where $P_\nu={\rm Diag.}(1,~1,~\lambda_2^*\rho^{m})$. The mass eigenstate
$|\psi_0\rangle=(0,~0,~1)^T$ of $S_{0p}$ with
zero eigenvalue goes to $|\psi_f\rangle\equiv U|\psi_0\rangle=V_\rho^\dagger|\psi_0\rangle = \lambda_2^*\rho^{m}
\left(0,~\frac{-1}{\sqrt{2}},~\frac{1}{\sqrt{2}}\right)^T$ and is independent of the unknown angle
$\theta_\nu$. Since
we are considering  inverted hierarchy, this state has to be associated with the third column of $U$. This leads to two
predictions namely, $\theta_{23}=\frac{\pi}{4}$ and $\theta_{13}=0$. The same eigenvector would describe the first
column of $U$ in case of the normal hierarchy and would lead to prediction $\theta_{12}$ or $\theta_{13}=\frac{\pi}{2}$
not realized by data. Thus the above example cannot accommodate normal hierarchy.

The above predictions are known to follow  if neutrino mass matrix ${\cal M}_{\nu f}$  in the flavor basis possesses
a $\mu$-$\tau$ symmetry. In the present case, ${\cal M}_{\nu f}$ is invariant under the symmetry $S_{\nu f}=U_l^\dagger
S_{0p}U_l$ with $U_l=V_\rho$ and $S_{0p}$ given by Eq. (\ref{sgm2n}). Requiring this invariance, one arrives at
\be \label{scalling-M}
{\cal M}_{\nu f}=\left(
\ba{ccc}  
X&A&A\\
A&B&B\\
A&B&B\\
\ea\right)~.\ee
This form is independent of the integer $m,~n,~p$ used in defining $S_{0p}$ and $T_{mn}$. It displays the scaling form  
with a massless neutrino which is studied in a number of papers \cite{Mohapatra:2006xy,*Blum:2007qm,*Joshipura:2009fu}.

The above form for neutrino mass matrix provides a good zeroth order approximation to realistic pattern with non-zero
$\theta_{13}$ for the following reason. In order to be able to do so, ${\cal M}_{\nu f}$ must be such that a small
perturbation to it can generate the correct mixing angles. It is found \cite{Gupta:2013it} that this does not happen in
case of an arbitrary $\mu$-$\tau$ symmetric neutrino mas matrices. Only those possessing inverted or quasi degenerate
spectrum can lead to correct mixing pattern with small perturbation. Since the above mass matrix implies inverted
hierarchy, small perturbations in $A$ and $B$ which can arise from the residual symmetry breaking are expected to 
generate the correct mixing pattern.

\subsection{Mixing pattern with $S_4(2)$}
\label{s42-section}
$S_4(2)$ is a member of the  group series  $S_4(m)\equiv A_4\rtimes Z_{2^m}$ which are subgroups of $U(3)$ admitting
a three dimensional faithful irreducible representation \cite{Ludl:2010bj}. Properties of $S_4(2)$ are studied by Ludl
\cite{Ludl:2010bj} and we use this group  here to show that it can be used to predict BM mixing in case of
the inverted hierarchy.

The first member of the series $S_4(1)$ is isomorphic to $S_4$. This group  has been used to predict both 
TBM \cite{Altarelli:2010gt,*Altarelli:2012ss,*King:2013eh} and BM
\cite{deAdelhartToorop:2011re,Altarelli:2009gn,*Toorop:2010yh,*Patel:2010hr,*Meloni:2011fx} mixing patterns. It is
useful to briefly recapitulate how this is achieved. The presentations of $S_4$ are  given in a three dimensional
representation by the matrices $A\equiv s_1$ and $B\equiv T$ defined in Eq. (\ref{snut}) and the $\mu$-$\tau$
interchange symmetry $C$:
$$ C=\left(
\ba{ccc}
1&0&0\\
0&0&1\\
0&1&0\\
\ea \right).$$
The TBM mixing arises when $S_4$ gets broken to $G_\nu=Z_2\times Z_2$ and $G_l=Z_3$ with $S_1=A$ and $S_2=C$
generating $Z_2\times Z_2$ and $T=B$ generating $Z_3$. The BM mixing is obtained with $G_\nu=Z_2\times Z_2$ and
$G_l=Z_4$.  The corresponding generators are give by: 
\beqa\label{bmgen}
S_1&\equiv& BAB^{-1}=\left(
\ba{ccc}
-1&0 &0 \\
0&-1 &0 \\
0&0 &1\\
\ea\right)~,\nonumber \\
S_2&\equiv& BCB^{-1}=\left(
\ba{ccc}
0&1 &0 \\
1&0 &0 \\
0&0 &1\\
\ea\right)~,\nonumber \\
T&\equiv& BAB^{-1}C=\left(
\ba{ccc}
-1&0 &0 \\
0&0 &-1 \\
0&1 &0\\
\ea\right)~.\eeqa

Let us now discuss $S_4(2)$. It contains $A_4$ subgroup which is given by $A$ and $B$ defined above in case of $S_4$. 
The $\mu$-$\tau$ symmetry operator $C$ is replaced by $C'\equiv i  C$ generating  a $Z_4$ group used in defining the
semi-direct product $A_4\rtimes Z_4$. $A$, $B$ and $C'$ provide a presentation of $S_4(2)$
\cite{Ludl:2010bj}. In order to
obtain a massless state at least one of the generator in $G_\nu$ has to be $Z_N$ with $N>2$. After a systematic search
over various  $Z_N\in S_4(2)$ we find that getting TBM mixing along with the right mass hierarchy is not possible but
one can obtain the BM mixing with inverted hierarchy. This is obtained when both $G_\nu$ and $G_l$ correspond to $Z_4$
symmetry with the generators $S_\nu$, $T_l\in S_4(2)$: 
\be \label{s42-ST}
S_\nu\equiv AB^{-1}C' =\left( \ba{ccc}
0&i&0\\
-i&0&0\\
0&0&-i\\
\ea \right)~~~{\rm and}~~~
T_l\equiv B A B^{-1}C'=\left( \ba{ccc}
-i&0&0\\
0&0&-i\\
0&i&0\\
\ea \right)~. \ee
Not that $S_\nu$ and $T_l$  are permutations of each other. $T_l$ is diagonalized by a rotation $U_l=P_2R_{23}(\pi/4)$
in the 2-3 plane and $S_\nu$ by $U_\nu=P_1R_{12}(\pi/4)$ where $P_2={\rm  Diag.}(1,~i,~1)$ and
$P_1={\rm Diag.}(i,~1,~1)$. As a result, one obtains the PMNS mixing matrix
\be \label{bm}
U=R_{23}^T(\pi/4)P_2^*P_1 R_{12}(\pi/4) \ee
with predictions $\theta_{12}=\theta_{23}=\frac{\pi}{4}$ and $\theta_{13}=0$.  The $S_\nu$ has the eigenvalues
$(1,~-1,~-i)$ and this choice describes inverted hierarchy with the eigenvector of a massless state in flavor space
$|\psi_f\rangle=\left(0,~\frac{1}{\sqrt{2}},~\frac{1}{\sqrt{2}}\right)^T$. The neutrino mass matrix in the flavor
basis invariant under the symmetry $S_{\nu f}=U_l^\dagger S_\nu U_l$ in this case is given by
\be
{\cal M}_{\nu f}=\left(
\ba{ccc}  
2B&A&A\\
A&B&B\\
A&B&B\\
\ea\right)~.\ee
This leads to a massless neutrino as expected. It also gives the BM mixing except in a special case
with Re($AB^*)=0$. One gets degenerate neutrino pair in this case and the solar angle is undefined in this limit.

One can obtain normal hierarchy by interchanging the role of $S_\nu$ and $T_l$ in Eq. (\ref{s42-ST}). However in this
case, the first and the  third columns of $U$ in Eq. (\ref{bm}) get interchanged and it predicts $\theta_{13}=\pi/4$
and $\theta_{12}=\pi/2$ which are far away from their experimental values. We also searched for alternative solutions
with normal hierarchy within $S_4(2)$. A solution which comes closest to the observed mixing pattern is given by  the
choice $S_\nu=BAB^{-1}C' \in Z_4$ and $T_l=B \in Z_3$. This predicts $\theta_{23}\approx$ 36.2\textdegree,
$\theta_{12}\approx$ 53.8\textdegree and $\theta_{13}\approx$ 12.2\textdegree. This solution accommodates non-zero
$\theta_{13}$ and non-maximal $\theta_{23}$ at the zeroth order approximation itself but it requires rather large
corrections in order to get viable solar mixing angle. In the next section, we provide a specific model realization of the above solution and discuss the suitable corrections which can lead to viable mixing angles.

\section{An $S_4(2)$ model of a massless neutrino}
\label{model}

The $S_4(2)$ group contains four 1-dimensional ($1_1$, $1_2$, $1_3$, $1_4$), two 2-dimensional ($2_1$, $2_2$) and
four 3-dimensional ($3_1$, $3_2$, $3_3$, $3_4$) irreducible representations (irreps) \cite{Ludl:2010bj}. The above set includes the irreps of $S_4$ which are $1_1$, $1_2$, $2_1$, $3_1$ and $3_2$. The remaining irreps of $S_4(2)$ can be obtained by multiplying each irrep of $S_4$ with 1-dimensional irreps of $Z_4$ (see the Appendix for more details). The structure
of tensor product decomposition can easily be obtained using those of $S_4$ as discussed in \cite{Ludl:2010bj} and in the Appendix here. We also give in the Appendix representation matrices in a chosen basis and multiplication rules relevant for
the model discussed here. Note that in our basis all the irreps of $S_4$ are real. Further, one can see that $1_3^*=1_4$ and $3_3^*=3_4$.

We now present an extension of the minimal suspersymmetric standard model (MSSM) based on $S_4(2)$
symmetry which can lead to a massless neutrino and bi-maximal mixing. Let's consider the left-handed
lepton doublet $L$ transforming as a triplet $3_3$ which does not belong to $S_4$. The right-handed
charged leptons are assigned to $e^c\sim 1_3$, $\mu^c\sim 1_2$ and $\tau^c\sim 1_1$ representations
of $S_4(2)$. We require five flavon fields $\phi_1^l$ ($\sim 3_2$), $\phi_2^l$ ($\sim 3_3$),
$\phi_3^l$ ($\sim 3_4$), $\phi^{\nu}$ ($\sim 3_2$) and $\chi^{\nu}$ ($\sim 2_1$) in order to break
$S_4(2)$ into $G_l=Z_4$ in the charged lepton sector and $G_\nu=Z_4$ in the neutrino sector as discussed in the previous section. The light neutrino masses arise through a dim-5 operator $L L H_u H_u$ where $H_u$ and $H_d$ are the
MSSM Higgs doublets which are singlets under $S_4(2)$. In order to distinguish between $\phi_1^l$ and
$\phi^\nu$, we impose an additional $Z_2$ symmetry under which $\phi_{1,2,3}^l$ and right-handed charged leptons are odd and the remaining fields are even. The $S_4(2)\times Z_2$ invariant Yukawa superpotential at the leading order can be written as
\beqa \label{suppot}
{\cal W}_Y &=& \left( y_e L\phi_1^l e^c + y_\mu L\phi_2^l \mu^c +
y_\tau L\phi_3^l \tau^c  \right) \frac{H_d}{\Lambda} \nonumber \\
&+& \left( y_1 LL \phi^{\nu}  + y_2 LL \chi^{\nu} \right)
\frac{H_u H_u}{\Lambda^2}~.
\eeqa

Let's now discuss the breaking of $S_4(2)$ symmetry. The bi-maximal mixing and inverted hierarchy can be achieved if $S_4(2)$ is broken to $G_l=Z_4$ in the charged lepton sector and $G_\nu=Z_4$ in the neutrino sector as already mentioned in Eq. (\ref{s42-ST}). In order to ensure that $S_4(2)$ breaks into $G_l$ in the charged lepton sector, the vevs of 
$\phi_{1,2,3}^l$ have to be invariant under $G_l$. This can be achieved by taking $T_l$ in the
representation corresponding to the flavon field $\phi_i^l$ and demanding $T_l\vev{\phi_i^l}=\vev{\phi_i^l}$.
For example, in case of $\phi_1^l$ ($\sim 3_2$) one gets  $T_l(3_2) \equiv  T_l (A\rightarrow A,~B\rightarrow B,~C'
\rightarrow i C') = B A B^{-1}(iC')$ and an invariance under $T_l(3_2)$ requires $\vev{\phi_1^l}=(\upsilon_{\phi_1},0,0)^T$.
Following the same strategy for $\phi_2^l$ and $\phi_3^l$, one obtains the vacuum structures
\be \label{vac-cl}
\vev{\phi_1^l}=\upsilon_{\phi_1}(1,0,0)^T,~~~\vev{\phi_2^l}=\upsilon_{\phi_2}(0,1,i)^T,~~{\rm
and}~~\vev{\phi_3^l}=\upsilon_{\phi_3}(0,1,-i)^T. \ee
We similarly find the vacuum structures of $\phi^\nu$ and $\chi^\nu$ invariant under $G_\nu$: 
\be \label{vac-nu}
\vev{\phi^\nu}=\upsilon_{\phi}(0,0,1)^T,~~{\rm and}~~\vev{\chi^\nu}=\upsilon_{\chi}(1,-\sqrt{3})^T. \ee

After the flavor symmetry is broken by the vevs of flavons as given in Eqs. (\ref{vac-cl}, \ref{vac-nu}) and after the
breaking of electroweak symmetry, we obtain the following mass matrices for charged leptons and
neutrinos :
\be \label{mass-matrices}
M_l = \upsilon_d \left( \ba{ccc}
y_e&0&0\\
0&y_\mu&y_\tau\\
0&i y_\mu&-i y_\tau\\
\ea \right)~~~{\rm and}~~~
M_\nu = \frac{\upsilon_u^2}{\Lambda} \left( \ba{ccc}
y_2&y_1&0\\
y_1&-y_2&0\\
0&0&0\\
\ea \right)~
, \ee
where $\upsilon_{u,d}=\vev{H_{u,d}}$ and the Yukawa couplings $y_i$ are suitably redefined by absorbing the flavon vevs. In the derivation of the above mass matrices from the Lagrangian (\ref{suppot}), we have used the multiplication rules listed in Eqs. (\ref{mr1}-\ref{mr4}) in the Appendix. As it is set by the vacuum structures of the flavon fields, the above mass matrices satisfy
\be \label{invs42}
S_\nu^T M_\nu S_\nu=M_\nu~~~{\rm and}~~~T_l^\dagger M_lM_l^\dagger T_l=M_lM_l^\dagger~\ee
and, at the leading order, lead to a massless neutrino with bimaximal mixing in the lepton sector.

The above predictions can be corrected and made viable in the following two possible ways. First, the next to leading order effects in general break both $G_l$ and $G_\nu$ generating corrections to the bi-maximal mixing pattern as well as generating a mass for the neutrino. However such corrections are generically assumed to be very small and they may not induce relatively large corrections required in the solar and reactor angles. Second, a breaking in the residual symmetry of the charged leptons only, namely in $G_l$, may arise at the leading order itself due to the presence of additional fields in the spectrum. In this scenario, the leading order prediction of a massless neutrino is not perturbed by such corrections. Here, we provide an example of the second type. Consider the presence of an additional $Z_2$ odd flavon field ${\phi'}_2^l$ which transforms as $3_3$ of $S_4(2)$. This adds a piece of interaction in Eq. (\ref{suppot}) proportional to $\epsilon L{\phi'}_2^l \mu^c$. If ${\phi'}_2^l$ takes vev in the direction $(1,0,0)^T$ which does not respect the $G_l$ symmetry characterized by $T_l$ in Eq. (\ref{s42-ST}). This leads to the following correction in $M_l$:
\be \label{corrected-ml}
M_l = \upsilon_d \left( \ba{ccc}
y_e&\epsilon&0\\
0&y_\mu&y_\tau\\
0&i y_\mu&-i y_\tau\\
\ea \right). \ee
The resulting $M_lM_l^\dagger$ takes the following form
\be \label{corrected-mlml}
M_l M_l^\dagger = \upsilon_d^2 \left( \ba{ccc}
x&a&-ia\\
a^*&y&-iz\\
ia^*&iz&y\\
\ea \right), \ee
where $x,~y$ and $z$ are real parameters. The group $Z_4$ breaks completely once the ${\phi'}_2^l$ acquires vev. However, the matrix $M_lM_l^\dagger$ now possess an accidental $Z_2$ symmetry generated by
\be \label{newtl}
T_l'=\left( \ba{ccc}
1&0&0\\
0&0&-i\\
0&i&0\\
\ea \right)~. \ee
Note that $T_l'$ is not even a subgroup of $S_4(2)$ since all generators of $S_4(2)$ in our basis are either purely real or purely imaginary. The $M_l M_l^\dagger$ in Eq. (\ref{corrected-mlml}) can be diagonalized by $U_l=P_3 R_{23}(\pi/4) U_{12}(\theta,\alpha)$ where $P_3 = {\rm Diag.}(1,1,i)$, 
\be \label{u12}
U_{12}(\theta,\alpha) = \left( \ba{ccc}
\cos\theta& -e^{i \alpha} \sin\theta &0\\
e^{-i \alpha} \sin\theta & \cos\theta&0\\
0 & 0& 1\\
\ea \right), \ee
where $\tan2\theta = 2\sqrt{2}|a|/(x-y-z)$ and $\alpha={\rm arg}(a)$. The neutrino mass matrix in Eq. (\ref{mass-matrices}) is diagonalized by $U_\nu=P_2 R_{12}(\pi/4)$ where $P_2={\rm Diag.}(1,i,1)$. The resulting $U_{PMNS}\equiv U= U_l^\dagger U_\nu$ predicts the following correlations among the leptonic mixing angles.
\beqa \label{predictions}
U_{e3} &=& -i e^{i \alpha} \frac{\sin\theta}{\sqrt{2}}, \nonumber \\
U_{e2} &=& -\frac{1}{\sqrt{2}}\left( \cos\theta - i e^{i \alpha} \frac{\sin\theta}{\sqrt{2}}\right)  , \nonumber \\
U_{\mu3} &=& -\frac{i}{\sqrt{2}}\cos\theta~. \eeqa
For $\alpha\approx -\pi/2$ and $\theta \approx 0.23$, it predicts $\sin^2\theta_{13}\approx0.026$, $\sin^2\theta_{12}\approx 0.339$ and $\sin^2\theta_{23}\approx0.487$ which are in agreement within the $3\sigma$ ranges of their global fit values. Interestingly, the angle $\theta$ is close to the Cabibbo angle. Also, the amount of perturbation $\epsilon$ required to correct the mixing angles is quite small
\be \label{epsilon} \frac{|\epsilon| \upsilon_d}{m_\tau}\approx
\frac{m_\mu}{m_\tau}\frac{\theta}{\sqrt{2}}\approx 10^{-2} . \ee
Having fixed  $\alpha$ and $\theta$ one now predicts $\delta\sim \pi$ for the Dirac CP phase implying the near absence of CP violation in neutrino oscillations. The above values of $\theta$ and $\alpha$ fix the parameter $a$ while the three charged lepton masses are determined using the remaining three free parameters in Eq. (\ref{corrected-mlml}). Similarly, the solar and atmospheric mass squared differences determine the complex parameters $y_1$ and $y_2$ in $M_\nu$ in Eq. (\ref{mass-matrices}). The above corrections generate viable mixing pattern at the leading order maintaining the prediction of a massless neutrino.

\section{Discussions}
We have addressed here  the problem of finding appropriate groups $G$ which can  lead to a massless neutrino by
identifying   residual symmetry  $G_\nu$ of the neutrino mass matrix  in this case.  Resulting $G_\nu$  is larger than
the conventional $Z_2\times Z_2$ groups used extensively in case of massive Majorana neutrinos. As argued here, the
groups  $G$ which contain  $G_\nu$  are subgroups of $U(3)$ rather than of  $SU(3)$ and hence are quite different from
the ones used so far in literature \cite{Altarelli:2010gt,*Altarelli:2012ss,*King:2013eh,Ishimori:2010au}. We have
considered  group series $\Sigma(3N^3)$, $\Sigma(2N^2)$ and the group $S_4(2)\equiv A_4\rtimes Z_4$ as possible
examples which contain $G_\nu$ implying a massless neutrino. It is shown that last two of these lead to good zeroth
order approximation to realistic mixing.  

Suitable perturbations are required in the discussed examples to generate the viable
leptonic mixing angles. Depending on the nature of corrections, the following possibilities may
arise: (1) If perturbation only breaks the symmetry group $G_l$ and keeps $G_\nu$ intact then we can
generate correction to mixing without generating a mass for a neutrino. An explicit example of this
case is presented in the section \ref{model} where we discuss a breaking of $G_l$ which can
lead to a realistic mixing pattern maintaining a massless neutrino. Similar argument works in the
case of $\Sigma(2N^2)$ where the solar angle is not fixed by the symmetry. (2) Even if perturbation
is such that $G_\nu$ is affected and massless neutrino picks a mass, we would have found a new
flavor symmetry to describe realistic masses and mixing which cannot be arrived at by insisting that
$G_\nu$ contains a $Z_2 \times Z_2$ symmetry. Clearly, these possibilities require separate
and detailed investigations of the specific models based on the symmetry groups proposed here.

The approach pursued here shares advantages and disadvantages of similar approach used extensively
\cite{Altarelli:2010gt,*Altarelli:2012ss,*King:2013eh,Ishimori:2010au} for relating leptonic mixing to symmetry in case
of massive neutrinos.  Advantage is that residual symmetries  $G_\nu$ and $G_l$ inferred from experimental knowledge
tell us what could be the underlying flavor symmetry groups which when broken to $G_\nu$ and $G_l$ lead to  definite
mixing pattern and in our case also a massless neutrino.  Disadvantage is that other groups $G'$ which may not contain
$G_\nu$ fully may also lead to the same mixing pattern as $G$ and the remaining symmetry in $G_\nu$ may arise
accidentally. Well-known examples are models based on $A_4$
\cite{Altarelli:2010gt,*Altarelli:2012ss,*King:2013eh,Ishimori:2010au} leading  to TBM mixing in spite of the fact
that $A_4$  contains only one of the two  $Z_2$ symmetries required to obtain TBM. The other one arises as accidental
symmetry in these models. Similarly, in our case also, a massless neutrino may result from an accidental symmetry of
the mass matrix. A simple example would be the type-I seesaw model with three active and two massive right handed 
neutrinos. In this model, a massless neutrino and hence the corresponding residual symmetry would arise purely for
`kinematical' reasons without imposing any flavor symmetry. But barring such cases, the approach studied here allows a
systematic way of identifying flavor groups leading to a massless neutrino and some definite mixing patterns.\\

\noindent{\bf Acknowledgments:} A.S.J. thanks the Department of Science and Technology, Government of India for support
under the J. C. Bose National Fellowship programme, grant no. SR/S2/JCB-31/2010. K.M.P. thanks the Theory Division of 
Physical Research Laboratory for hospitality and support.

\begin{appendix}
\section{The group $S_4(2)$}
The group $S_4(2)$ is a subgroup of $U(3)$ and it has a structure equivalent to $A_4\rtimes Z_4$. The properties of 
$S_4(2)$ are studied in \cite{Ludl:2010bj} and we use many of the results obtained there. In the basis considered in the text, the generators of a faithful 3-dimensional irrep of $S_4(2)$ are given by
\be \label{s42-generators}
A = \left( \ba{ccc}
1&0&0\\
0&-1&0\\
0&0&-1\\
\ea \right),~~~
B = \left( \ba{ccc}
0&1&0\\
0&0&1\\
1&0&0\\
\ea \right)~~~{\rm and}~~~
C' =i\left( \ba{ccc}
1&0&0\\
0&0&1\\
0&1&0\\
\ea \right)~
.\ee

The irreps of $S_4(2)$ in the above basis are given by \cite{Ludl:2010bj}
\beqa \label{s42-irreps}
1_1 &:& A=1,~B=1,~C'=1; \nonumber \\
1_2 &:& A=1,~B=1,~C'=-1; \nonumber \\
1_3 &:& A=1,~B=1,~C'=i; \nonumber \\
1_4 &:& A=1,~B=1,~C'=-i; \nonumber \\
2_1 &:& A=\left( \ba{cc} 1&0\\ 0&1 \\\ea \right),~B=\frac{1}{2}\left( \ba{cc} -1&\sqrt{3}\\ -\sqrt{3}&-1 \\\ea
\right),~C'=\left( \ba{cc} 1&0\\ 0&-1 \\\ea \right); \nonumber \\
2_2 &:& A=\left( \ba{cc} 1&0\\ 0&1 \\\ea \right),~B=\frac{1}{2}\left( \ba{cc} -1&\sqrt{3}\\ -\sqrt{3}&-1 \\\ea
\right),~C'=i\left( \ba{cc} 1&0\\ 0&-1 \\\ea \right); \nonumber \\
3_1 &:& A=A,~B=B,~C'=-iC'; \nonumber \\
3_2 &:& A=A,~B=B,~C'=iC'; \nonumber \\
3_3 &:& A=A,~B=B,~C'=C'; \nonumber \\
3_4 &:& A=A,~B=B,~C'=-C'. \eeqa
All the irreps of $S_4$ ($1_1$, $1_2$, $2_1$, $3_1$, $3_2$) are irreps of $S_4(2)$ too as well as all the irreps of
$Z_4$ ($1_1$, $1_2$, $1_3$, $1_4$) are irreps of $S_4(2)$. Further, one can obtain all the irreps of $S_4(2)$ from the
irreps of $S_4$ by multiplying them with the irreps of $Z_4$. For example, one gets
\be \label{s4-s42}
2_1 \equiv 2_1 \otimes 1_3,~3_3 \equiv 3_1 \otimes 1_3,~3_4 \equiv 3_1 \otimes 1_4 \ee

\subsection{Tensor products}
The tensor products for irreps of $S_4$ are given as:
\beqa \label{tps4}
1_1 \otimes \chi_i &=& \chi_i; \nonumber \\
1_2 \otimes 1_2 &=& 1_1;~1_2\otimes2_1=2_1;~1_2\otimes3_1=3_2;~1_2\otimes3_2=3_1; \nonumber \\
2_1 \otimes 2_1 &=& 1_1 \oplus 1_2 \oplus 2_1;~2_1\otimes3_1=2_1\otimes3_2=3_1\oplus3_2; \nonumber \\
3_1 \otimes 3_1 &=&3_2 \otimes 3_2 = 1_1 \oplus 2_1 \oplus 3_1 \oplus 3_2; \nonumber \\
3_1 \otimes 3_2 &=& 1_2 \oplus 2_1 \oplus 3_1 \oplus 3_2 \eeqa

The tensor products of remaining irreps of $S_4(2)$ can be obtained in a straightforward way by
multiplying the irreps of $S_4$ with suitable 1-dimensional irrep of $Z_4$. For example, one obtains using Eqs.
(\ref{s4-s42})
\beqa \label{tps42-1}
1_3 \otimes 1_3 &=& 1_4 \otimes 1_4= 1_2;~1_3 \otimes 1_4 = 1_1; \nonumber \\
3_1 \otimes 3_3 &=& 3_1 \otimes (3_1 \otimes 1_3) = 1_3 \oplus 2_2 \oplus 3_3 \oplus 3_4; \nonumber \\
3_2 \otimes 3_3 &=& 3_2 \otimes (3_1 \otimes 1_3) = 1_4 \oplus 2_2 \oplus 3_3 \oplus 3_4; \nonumber \\
3_3 \otimes 3_3 &=& (3_1 \otimes 1_3) \otimes (3_1 \otimes 1_3) = 1_2 \oplus 2_1 \oplus 3_1 \oplus 3_2; \nonumber \\
3_3 \otimes 3_4 &=& (3_1 \otimes 1_3) \otimes (3_1 \otimes 1_4) = 1_1 \oplus 2_1 \oplus 3_1 \oplus 3_2\eeqa

\subsection{Multiplication rules}
The multiplication rules for tensor products in the basis we have chosen are as the below. Here, we give only those multiplication rules relevant for the model presented in the text.

\beqa \label{mr1}
(x)_{1_i} \otimes (y)_{1_j} &=& (xy)_{1_k}~. \eeqa

\beqa \label{mr2}
\left(\ba{c}
x_1\\ x_2  \ea \right)_{2_i} \otimes \left(\ba{c}
y_1\\ y_2  \ea \right)_{2_j}
&=& \frac{1}{\sqrt{2}}(x_1y_1+x_2y_2)_{1_k} \oplus ...~.  \eeqa

\beqa \label{mr3}
\left(\ba{c}
x_1\\ x_2 \\ x_3 \ea \right)_{3_i} \otimes \left(\ba{c}
y_1\\ y_2 \\ y_3 \ea \right)_{3_j}
&=& \frac{1}{\sqrt{3}}(x_1y_1+x_2y_2+x_3y_3)_{1_k} \oplus ...~.  \eeqa

\beqa \label{mr4}
\left(\ba{c}
x_1\\ x_2 \\ x_3 \ea \right)_{3_3} \otimes \left(\ba{c}
y_1\\ y_2 \\ y_3 \ea \right)_{3_3}
&=& \frac{1}{\sqrt{3}}(x_1y_1+x_2y_2+x_3y_3)_{1_2} \nonumber \\
&\oplus&\left(\ba{c}
\frac{1}{\sqrt{2}}(x_2y_2-x_3y_3)\\ \frac{1}{\sqrt{6}}(2x_1y_1-x_2y_2-x_3y_3) \ea \right)_{2_1} \nonumber \\
&\oplus&\frac{1}{\sqrt{2}}\left(\ba{c}
x_2y_3-x_3y_1\\ x_3y_1-x_1y_3\\ x_1y_2-x_2y_1\\ \ea
\right)_{3_1} \nonumber \\ 
&\oplus&\frac{1}{\sqrt{2}}\left(\ba{c}
x_2y_3+x_3y_1\\ x_3y_1+x_1y_3\\ x_1y_2+x_2y_1\\ \ea
\right)_{3_2} \eeqa	

\end{appendix}

\bibliographystyle{apsrev4-1}
\bibliography{refs-massless}

\begin{thebibliography}{39}%
\makeatletter
\providecommand \@ifxundefined [1]{%
 \@ifx{#1\undefined}
}%
\providecommand \@ifnum [1]{%
 \ifnum #1\expandafter \@firstoftwo
 \else \expandafter \@secondoftwo
 \fi
}%
\providecommand \@ifx [1]{%
 \ifx #1\expandafter \@firstoftwo
 \else \expandafter \@secondoftwo
 \fi
}%
\providecommand \natexlab [1]{#1}%
\providecommand \enquote  [1]{``#1''}%
\providecommand \bibnamefont  [1]{#1}%
\providecommand \bibfnamefont [1]{#1}%
\providecommand \citenamefont [1]{#1}%
\providecommand \href@noop [0]{\@secondoftwo}%
\providecommand \href [0]{\begingroup \@sanitize@url \@href}%
\providecommand \@href[1]{\@@startlink{#1}\@@href}%
\providecommand \@@href[1]{\endgroup#1\@@endlink}%
\providecommand \@sanitize@url [0]{\catcode `\\12\catcode `\$12\catcode
  `\&12\catcode `\#12\catcode `\^12\catcode `\_12\catcode `\%12\relax}%
\providecommand \@@startlink[1]{}%
\providecommand \@@endlink[0]{}%
\providecommand \url  [0]{\begingroup\@sanitize@url \@url }%
\providecommand \@url [1]{\endgroup\@href {#1}{\urlprefix }}%
\providecommand \urlprefix  [0]{URL }%
\providecommand \Eprint [0]{\href }%
\providecommand \doibase [0]{http://dx.doi.org/}%
\providecommand \selectlanguage [0]{\@gobble}%
\providecommand \bibinfo  [0]{\@secondoftwo}%
\providecommand \bibfield  [0]{\@secondoftwo}%
\providecommand \translation [1]{[#1]}%
\providecommand \BibitemOpen [0]{}%
\providecommand \bibitemStop [0]{}%
\providecommand \bibitemNoStop [0]{.\EOS\space}%
\providecommand \EOS [0]{\spacefactor3000\relax}%
\providecommand \BibitemShut  [1]{\csname bibitem#1\endcsname}%
\let\auto@bib@innerbib\@empty
%</preamble>
\bibitem [{\citenamefont {Gonzalez-Garcia}\ \emph {et~al.}(2012)\citenamefont
  {Gonzalez-Garcia}, \citenamefont {Maltoni}, \citenamefont {Salvado},\ and\
  \citenamefont {Schwetz}}]{GonzalezGarcia:2012sz}%
  \BibitemOpen
  \bibfield  {author} {\bibinfo {author} {\bibfnamefont {M.}~\bibnamefont
  {Gonzalez-Garcia}}, \bibinfo {author} {\bibfnamefont {M.}~\bibnamefont
  {Maltoni}}, \bibinfo {author} {\bibfnamefont {J.}~\bibnamefont {Salvado}}, \
  and\ \bibinfo {author} {\bibfnamefont {T.}~\bibnamefont {Schwetz}},\ }\href
  {\doibase 10.1007/JHEP12(2012)123} {\bibfield  {journal} {\bibinfo  {journal}
  {JHEP}\ }\textbf {\bibinfo {volume} {1212}},\ \bibinfo {pages} {123}
  (\bibinfo {year} {2012})},\ \Eprint {http://arxiv.org/abs/1209.3023}
  {arXiv:1209.3023 [hep-ph]} \BibitemShut {NoStop}%
%%CITATION = ARXIV:1209.3023;%%
\bibitem [{\citenamefont {Fogli}\ \emph {et~al.}(2012)\citenamefont {Fogli},
  \citenamefont {Lisi}, \citenamefont {Marrone}, \citenamefont {Montanino},
  \citenamefont {Palazzo} \emph {et~al.}}]{Fogli:2012ua}%
  \BibitemOpen
  \bibfield  {author} {\bibinfo {author} {\bibfnamefont {G.}~\bibnamefont
  {Fogli}}, \bibinfo {author} {\bibfnamefont {E.}~\bibnamefont {Lisi}},
  \bibinfo {author} {\bibfnamefont {A.}~\bibnamefont {Marrone}}, \bibinfo
  {author} {\bibfnamefont {D.}~\bibnamefont {Montanino}}, \bibinfo {author}
  {\bibfnamefont {A.}~\bibnamefont {Palazzo}},  \emph {et~al.},\ }\href
  {\doibase 10.1103/PhysRevD.86.013012} {\bibfield  {journal} {\bibinfo
  {journal} {Phys.Rev.}\ }\textbf {\bibinfo {volume} {D86}},\ \bibinfo {pages}
  {013012} (\bibinfo {year} {2012})},\ \Eprint {http://arxiv.org/abs/1205.5254}
  {arXiv:1205.5254 [hep-ph]} \BibitemShut {NoStop}%
%%CITATION = ARXIV:1205.5254;%%
\bibitem [{\citenamefont {Forero}\ \emph {et~al.}(2012)\citenamefont {Forero},
  \citenamefont {Tortola},\ and\ \citenamefont {Valle}}]{Tortola:2012te}%
  \BibitemOpen
  \bibfield  {author} {\bibinfo {author} {\bibfnamefont {D.}~\bibnamefont
  {Forero}}, \bibinfo {author} {\bibfnamefont {M.}~\bibnamefont {Tortola}}, \
  and\ \bibinfo {author} {\bibfnamefont {J.}~\bibnamefont {Valle}},\ }\href
  {\doibase 10.1103/PhysRevD.86.073012} {\bibfield  {journal} {\bibinfo
  {journal} {Phys.Rev.}\ }\textbf {\bibinfo {volume} {D86}},\ \bibinfo {pages}
  {073012} (\bibinfo {year} {2012})},\ \Eprint {http://arxiv.org/abs/1205.4018}
  {arXiv:1205.4018 [hep-ph]} \BibitemShut {NoStop}%
%%CITATION = ARXIV:1205.4018;%%
\bibitem [{\citenamefont {Altarelli}\ and\ \citenamefont
  {Feruglio}(2010)}]{Altarelli:2010gt}%
  \BibitemOpen
  \bibfield  {author} {\bibinfo {author} {\bibfnamefont {G.}~\bibnamefont
  {Altarelli}}\ and\ \bibinfo {author} {\bibfnamefont {F.}~\bibnamefont
  {Feruglio}},\ }\href {\doibase 10.1103/RevModPhys.82.2701} {\bibfield
  {journal} {\bibinfo  {journal} {Rev.Mod.Phys.}\ }\textbf {\bibinfo {volume}
  {82}},\ \bibinfo {pages} {2701} (\bibinfo {year} {2010})},\ \Eprint
  {http://arxiv.org/abs/1002.0211} {arXiv:1002.0211 [hep-ph]} \BibitemShut
  {NoStop}%
%%CITATION = ARXIV:1002.0211;%%
\bibitem [{\citenamefont {Altarelli}\ \emph {et~al.}(2012)\citenamefont
  {Altarelli}, \citenamefont {Feruglio},\ and\ \citenamefont
  {Merlo}}]{Altarelli:2012ss}%
  \BibitemOpen
  \bibfield  {author} {\bibinfo {author} {\bibfnamefont {G.}~\bibnamefont
  {Altarelli}}, \bibinfo {author} {\bibfnamefont {F.}~\bibnamefont {Feruglio}},
  \ and\ \bibinfo {author} {\bibfnamefont {L.}~\bibnamefont {Merlo}},\
  }\href@noop {} {\  (\bibinfo {year} {2012})},\ \Eprint
  {http://arxiv.org/abs/1205.5133} {arXiv:1205.5133 [hep-ph]} \BibitemShut
  {NoStop}%
%%CITATION = ARXIV:1205.5133;%%
\bibitem [{\citenamefont {King}\ and\ \citenamefont
  {Luhn}(2013)}]{King:2013eh}%
  \BibitemOpen
  \bibfield  {author} {\bibinfo {author} {\bibfnamefont {S.~F.}\ \bibnamefont
  {King}}\ and\ \bibinfo {author} {\bibfnamefont {C.}~\bibnamefont {Luhn}},\
  }\href {\doibase 10.1088/0034-4885/76/5/056201} {\bibfield  {journal}
  {\bibinfo  {journal} {Rept.Prog.Phys.}\ }\textbf {\bibinfo {volume} {76}},\
  \bibinfo {pages} {056201} (\bibinfo {year} {2013})},\ \Eprint
  {http://arxiv.org/abs/1301.1340} {arXiv:1301.1340 [hep-ph]} \BibitemShut
  {NoStop}%
%%CITATION = ARXIV:1301.1340;%%
\bibitem [{\citenamefont {Ishimori}\ \emph {et~al.}(2010)\citenamefont
  {Ishimori}, \citenamefont {Kobayashi}, \citenamefont {Ohki}, \citenamefont
  {Shimizu}, \citenamefont {Okada} \emph {et~al.}}]{Ishimori:2010au}%
  \BibitemOpen
  \bibfield  {author} {\bibinfo {author} {\bibfnamefont {H.}~\bibnamefont
  {Ishimori}}, \bibinfo {author} {\bibfnamefont {T.}~\bibnamefont {Kobayashi}},
  \bibinfo {author} {\bibfnamefont {H.}~\bibnamefont {Ohki}}, \bibinfo {author}
  {\bibfnamefont {Y.}~\bibnamefont {Shimizu}}, \bibinfo {author} {\bibfnamefont
  {H.}~\bibnamefont {Okada}},  \emph {et~al.},\ }\href {\doibase
  10.1143/PTPS.183.1} {\bibfield  {journal} {\bibinfo  {journal}
  {Prog.Theor.Phys.Suppl.}\ }\textbf {\bibinfo {volume} {183}},\ \bibinfo
  {pages} {1} (\bibinfo {year} {2010})},\ \Eprint
  {http://arxiv.org/abs/1003.3552} {arXiv:1003.3552 [hep-th]} \BibitemShut
  {NoStop}%
%%CITATION = ARXIV:1003.3552;%%
\bibitem [{\citenamefont {Lam}(2007)}]{Lam:2007qc}%
  \BibitemOpen
  \bibfield  {author} {\bibinfo {author} {\bibfnamefont {C.}~\bibnamefont
  {Lam}},\ }\href {\doibase 10.1016/j.physletb.2007.09.032} {\bibfield
  {journal} {\bibinfo  {journal} {Phys.Lett.}\ }\textbf {\bibinfo {volume}
  {B656}},\ \bibinfo {pages} {193} (\bibinfo {year} {2007})},\ \Eprint
  {http://arxiv.org/abs/0708.3665} {arXiv:0708.3665 [hep-ph]} \BibitemShut
  {NoStop}%
%%CITATION = ARXIV:0708.3665;%%
\bibitem [{\citenamefont {Lam}(2008{\natexlab{a}})}]{Lam:2008rs}%
  \BibitemOpen
  \bibfield  {author} {\bibinfo {author} {\bibfnamefont {C.}~\bibnamefont
  {Lam}},\ }\href {\doibase 10.1103/PhysRevLett.101.121602} {\bibfield
  {journal} {\bibinfo  {journal} {Phys.Rev.Lett.}\ }\textbf {\bibinfo {volume}
  {101}},\ \bibinfo {pages} {121602} (\bibinfo {year} {2008}{\natexlab{a}})},\
  \Eprint {http://arxiv.org/abs/0804.2622} {arXiv:0804.2622 [hep-ph]}
  \BibitemShut {NoStop}%
%%CITATION = ARXIV:0804.2622;%%
\bibitem [{\citenamefont {Lam}(2008{\natexlab{b}})}]{Lam:2008sh}%
  \BibitemOpen
  \bibfield  {author} {\bibinfo {author} {\bibfnamefont {C.}~\bibnamefont
  {Lam}},\ }\href {\doibase 10.1103/PhysRevD.78.073015} {\bibfield  {journal}
  {\bibinfo  {journal} {Phys.Rev.}\ }\textbf {\bibinfo {volume} {D78}},\
  \bibinfo {pages} {073015} (\bibinfo {year} {2008}{\natexlab{b}})},\ \Eprint
  {http://arxiv.org/abs/0809.1185} {arXiv:0809.1185 [hep-ph]} \BibitemShut
  {NoStop}%
%%CITATION = ARXIV:0809.1185;%%
\bibitem [{\citenamefont {Grimus}\ \emph {et~al.}(2009)\citenamefont {Grimus},
  \citenamefont {Lavoura},\ and\ \citenamefont {Ludl}}]{Grimus:2009pg}%
  \BibitemOpen
  \bibfield  {author} {\bibinfo {author} {\bibfnamefont {W.}~\bibnamefont
  {Grimus}}, \bibinfo {author} {\bibfnamefont {L.}~\bibnamefont {Lavoura}}, \
  and\ \bibinfo {author} {\bibfnamefont {P.}~\bibnamefont {Ludl}},\ }\href
  {\doibase 10.1088/0954-3899/36/11/115007} {\bibfield  {journal} {\bibinfo
  {journal} {J.Phys.}\ }\textbf {\bibinfo {volume} {G36}},\ \bibinfo {pages}
  {115007} (\bibinfo {year} {2009})},\ \Eprint {http://arxiv.org/abs/0906.2689}
  {arXiv:0906.2689 [hep-ph]} \BibitemShut {NoStop}%
%%CITATION = ARXIV:0906.2689;%%
\bibitem [{\citenamefont {Harrison}\ \emph {et~al.}(2002)\citenamefont
  {Harrison}, \citenamefont {Perkins},\ and\ \citenamefont
  {Scott}}]{Harrison:2002er}%
  \BibitemOpen
  \bibfield  {author} {\bibinfo {author} {\bibfnamefont {P.}~\bibnamefont
  {Harrison}}, \bibinfo {author} {\bibfnamefont {D.}~\bibnamefont {Perkins}}, \
  and\ \bibinfo {author} {\bibfnamefont {W.}~\bibnamefont {Scott}},\ }\href
  {\doibase 10.1016/S0370-2693(02)01336-9} {\bibfield  {journal} {\bibinfo
  {journal} {Phys.Lett.}\ }\textbf {\bibinfo {volume} {B530}},\ \bibinfo
  {pages} {167} (\bibinfo {year} {2002})},\ \Eprint
  {http://arxiv.org/abs/hep-ph/0202074} {arXiv:hep-ph/0202074 [hep-ph]}
  \BibitemShut {NoStop}%
%%CITATION = HEP-PH/0202074;%%
\bibitem [{\citenamefont {Xing}(2002)}]{Xing:2002sw}%
  \BibitemOpen
  \bibfield  {author} {\bibinfo {author} {\bibfnamefont {Z.-z.}\ \bibnamefont
  {Xing}},\ }\href {\doibase 10.1016/S0370-2693(02)01649-0} {\bibfield
  {journal} {\bibinfo  {journal} {Phys.Lett.}\ }\textbf {\bibinfo {volume}
  {B533}},\ \bibinfo {pages} {85} (\bibinfo {year} {2002})},\ \Eprint
  {http://arxiv.org/abs/hep-ph/0204049} {arXiv:hep-ph/0204049 [hep-ph]}
  \BibitemShut {NoStop}%
%%CITATION = HEP-PH/0204049;%%
\bibitem [{\citenamefont {Hernandez}\ and\ \citenamefont
  {Smirnov}(2013)}]{Hernandez:2013vya}%
  \BibitemOpen
  \bibfield  {author} {\bibinfo {author} {\bibfnamefont {D.}~\bibnamefont
  {Hernandez}}\ and\ \bibinfo {author} {\bibfnamefont {A.~Y.}\ \bibnamefont
  {Smirnov}},\ }\href@noop {} {\  (\bibinfo {year} {2013})},\ \Eprint
  {http://arxiv.org/abs/1304.7738} {arXiv:1304.7738 [hep-ph]} \BibitemShut
  {NoStop}%
%%CITATION = ARXIV:1304.7738;%%
\bibitem [{\citenamefont {Toorop}\ \emph {et~al.}(2011)\citenamefont {Toorop},
  \citenamefont {Feruglio},\ and\ \citenamefont {Hagedorn}}]{Toorop:2011jn}%
  \BibitemOpen
  \bibfield  {author} {\bibinfo {author} {\bibfnamefont {R.~d.~A.}\
  \bibnamefont {Toorop}}, \bibinfo {author} {\bibfnamefont {F.}~\bibnamefont
  {Feruglio}}, \ and\ \bibinfo {author} {\bibfnamefont {C.}~\bibnamefont
  {Hagedorn}},\ }\href {\doibase 10.1016/j.physletb.2011.08.013} {\bibfield
  {journal} {\bibinfo  {journal} {Phys.Lett.}\ }\textbf {\bibinfo {volume}
  {B703}},\ \bibinfo {pages} {447} (\bibinfo {year} {2011})},\ \Eprint
  {http://arxiv.org/abs/1107.3486} {arXiv:1107.3486 [hep-ph]} \BibitemShut
  {NoStop}%
%%CITATION = ARXIV:1107.3486;%%
\bibitem [{\citenamefont {Hernandez}\ and\ \citenamefont
  {Smirnov}(2012{\natexlab{a}})}]{Hernandez:2012ra}%
  \BibitemOpen
  \bibfield  {author} {\bibinfo {author} {\bibfnamefont {D.}~\bibnamefont
  {Hernandez}}\ and\ \bibinfo {author} {\bibfnamefont {A.~Y.}\ \bibnamefont
  {Smirnov}},\ }\href {\doibase 10.1103/PhysRevD.86.053014} {\bibfield
  {journal} {\bibinfo  {journal} {Phys.Rev.}\ }\textbf {\bibinfo {volume}
  {D86}},\ \bibinfo {pages} {053014} (\bibinfo {year} {2012}{\natexlab{a}})},\
  \Eprint {http://arxiv.org/abs/1204.0445} {arXiv:1204.0445 [hep-ph]}
  \BibitemShut {NoStop}%
%%CITATION = ARXIV:1204.0445;%%
\bibitem [{\citenamefont {Hernandez}\ and\ \citenamefont
  {Smirnov}(2012{\natexlab{b}})}]{Hernandez:2012sk}%
  \BibitemOpen
  \bibfield  {author} {\bibinfo {author} {\bibfnamefont {D.}~\bibnamefont
  {Hernandez}}\ and\ \bibinfo {author} {\bibfnamefont {A.~Y.}\ \bibnamefont
  {Smirnov}},\ }\href {\doibase 10.1103/PhysRevD.87.053005} {\bibfield
  {journal} {\bibinfo  {journal} {Phys.Rev.}\ }\textbf {\bibinfo {volume}
  {D87}},\ \bibinfo {pages} {053005} (\bibinfo {year} {2012}{\natexlab{b}})},\
  \Eprint {http://arxiv.org/abs/1212.2149} {arXiv:1212.2149 [hep-ph]}
  \BibitemShut {NoStop}%
%%CITATION = ARXIV:1212.2149;%%
\bibitem [{\citenamefont {Holthausen}\ \emph {et~al.}(2013)\citenamefont
  {Holthausen}, \citenamefont {Lim},\ and\ \citenamefont
  {Lindner}}]{Holthausen:2012wt}%
  \BibitemOpen
  \bibfield  {author} {\bibinfo {author} {\bibfnamefont {M.}~\bibnamefont
  {Holthausen}}, \bibinfo {author} {\bibfnamefont {K.~S.}\ \bibnamefont {Lim}},
  \ and\ \bibinfo {author} {\bibfnamefont {M.}~\bibnamefont {Lindner}},\ }\href
  {\doibase 10.1016/j.physletb.2013.02.047} {\bibfield  {journal} {\bibinfo
  {journal} {Phys.Lett.}\ }\textbf {\bibinfo {volume} {B721}},\ \bibinfo
  {pages} {61} (\bibinfo {year} {2013})},\ \Eprint
  {http://arxiv.org/abs/1212.2411} {arXiv:1212.2411 [hep-ph]} \BibitemShut
  {NoStop}%
%%CITATION = ARXIV:1212.2411;%%
\bibitem [{\citenamefont {de~Adelhart~Toorop}\ \emph
  {et~al.}(2012)\citenamefont {de~Adelhart~Toorop}, \citenamefont {Feruglio},\
  and\ \citenamefont {Hagedorn}}]{deAdelhartToorop:2011re}%
  \BibitemOpen
  \bibfield  {author} {\bibinfo {author} {\bibfnamefont {R.}~\bibnamefont
  {de~Adelhart~Toorop}}, \bibinfo {author} {\bibfnamefont {F.}~\bibnamefont
  {Feruglio}}, \ and\ \bibinfo {author} {\bibfnamefont {C.}~\bibnamefont
  {Hagedorn}},\ }\href {\doibase 10.1016/j.nuclphysb.2012.01.017} {\bibfield
  {journal} {\bibinfo  {journal} {Nucl.Phys.}\ }\textbf {\bibinfo {volume}
  {B858}},\ \bibinfo {pages} {437} (\bibinfo {year} {2012})},\ \Eprint
  {http://arxiv.org/abs/1112.1340} {arXiv:1112.1340 [hep-ph]} \BibitemShut
  {NoStop}%
%%CITATION = ARXIV:1112.1340;%%
\bibitem [{\citenamefont {Lam}(2013{\natexlab{a}})}]{Lam:2013ng}%
  \BibitemOpen
  \bibfield  {author} {\bibinfo {author} {\bibfnamefont {C.}~\bibnamefont
  {Lam}},\ }\href@noop {} {\  (\bibinfo {year} {2013}{\natexlab{a}})},\ \Eprint
  {http://arxiv.org/abs/1301.1736} {arXiv:1301.1736 [hep-ph]} \BibitemShut
  {NoStop}%
%%CITATION = ARXIV:1301.1736;%%
\bibitem [{\citenamefont {King}\ \emph {et~al.}(2013)\citenamefont {King},
  \citenamefont {Neder},\ and\ \citenamefont {Stuart}}]{King:2013vna}%
  \BibitemOpen
  \bibfield  {author} {\bibinfo {author} {\bibfnamefont {S.~F.}\ \bibnamefont
  {King}}, \bibinfo {author} {\bibfnamefont {T.}~\bibnamefont {Neder}}, \ and\
  \bibinfo {author} {\bibfnamefont {A.~J.}\ \bibnamefont {Stuart}},\
  }\href@noop {} {\  (\bibinfo {year} {2013})},\ \Eprint
  {http://arxiv.org/abs/1305.3200} {arXiv:1305.3200 [hep-ph]} \BibitemShut
  {NoStop}%
%%CITATION = ARXIV:1305.3200;%%
\bibitem [{\citenamefont {Parattu}\ and\ \citenamefont
  {Wingerter}(2011)}]{Parattu:2010cy}%
  \BibitemOpen
  \bibfield  {author} {\bibinfo {author} {\bibfnamefont {K.~M.}\ \bibnamefont
  {Parattu}}\ and\ \bibinfo {author} {\bibfnamefont {A.}~\bibnamefont
  {Wingerter}},\ }\href {\doibase 10.1103/PhysRevD.84.013011} {\bibfield
  {journal} {\bibinfo  {journal} {Phys.Rev.}\ }\textbf {\bibinfo {volume}
  {D84}},\ \bibinfo {pages} {013011} (\bibinfo {year} {2011})},\ \Eprint
  {http://arxiv.org/abs/1012.2842} {arXiv:1012.2842 [hep-ph]} \BibitemShut
  {NoStop}%
%%CITATION = ARXIV:1012.2842;%%
\bibitem [{\citenamefont {Hu}(2013)}]{Hu:2012ei}%
  \BibitemOpen
  \bibfield  {author} {\bibinfo {author} {\bibfnamefont {B.}~\bibnamefont
  {Hu}},\ }\href {\doibase 10.1103/PhysRevD.87.033002} {\bibfield  {journal}
  {\bibinfo  {journal} {Phys.Rev.}\ }\textbf {\bibinfo {volume} {D87}},\
  \bibinfo {pages} {033002} (\bibinfo {year} {2013})},\ \Eprint
  {http://arxiv.org/abs/1212.2819} {arXiv:1212.2819 [hep-ph]} \BibitemShut
  {NoStop}%
%%CITATION = ARXIV:1212.2819;%%
\bibitem [{\citenamefont {Lam}(2013{\natexlab{b}})}]{Lam:2013xs}%
  \BibitemOpen
  \bibfield  {author} {\bibinfo {author} {\bibfnamefont {C.}~\bibnamefont
  {Lam}},\ }\href@noop {} {\  (\bibinfo {year} {2013}{\natexlab{b}})},\ \Eprint
  {http://arxiv.org/abs/1301.3121} {arXiv:1301.3121 [hep-ph]} \BibitemShut
  {NoStop}%
%%CITATION = ARXIV:1301.3121;%%
\bibitem [{\citenamefont {Grimus}(2013)}]{Grimus:2013rw}%
  \BibitemOpen
  \bibfield  {author} {\bibinfo {author} {\bibfnamefont {W.}~\bibnamefont
  {Grimus}},\ }\href@noop {} {\  (\bibinfo {year} {2013})},\ \Eprint
  {http://arxiv.org/abs/1301.0495} {arXiv:1301.0495 [hep-ph]} \BibitemShut
  {NoStop}%
%%CITATION = ARXIV:1301.0495;%%
\bibitem [{\citenamefont {Grimus}\ and\ \citenamefont
  {Ludl}(2012)}]{Grimus:2011fk}%
  \BibitemOpen
  \bibfield  {author} {\bibinfo {author} {\bibfnamefont {W.}~\bibnamefont
  {Grimus}}\ and\ \bibinfo {author} {\bibfnamefont {P.~O.}\ \bibnamefont
  {Ludl}},\ }\href {\doibase 10.1088/1751-8113/45/23/233001} {\bibfield
  {journal} {\bibinfo  {journal} {J.Phys.}\ }\textbf {\bibinfo {volume}
  {A45}},\ \bibinfo {pages} {233001} (\bibinfo {year} {2012})},\ \Eprint
  {http://arxiv.org/abs/1110.6376} {arXiv:1110.6376 [hep-ph]} \BibitemShut
  {NoStop}%
%%CITATION = ARXIV:1110.6376;%%
\bibitem [{\citenamefont {Ludl}(2010)}]{Ludl:2010bj}%
  \BibitemOpen
  \bibfield  {author} {\bibinfo {author} {\bibfnamefont {P.~O.}\ \bibnamefont
  {Ludl}},\ }\href {\doibase 10.1088/1751-8113/44/13/139501,
  10.1088/1751-8113/43/39/395204} {\bibfield  {journal} {\bibinfo  {journal}
  {J.Phys.}\ }\textbf {\bibinfo {volume} {A43}},\ \bibinfo {pages} {395204}
  (\bibinfo {year} {2010})},\ \Eprint {http://arxiv.org/abs/1006.1479}
  {arXiv:1006.1479 [math-ph]} \BibitemShut {NoStop}%
%%CITATION = ARXIV:1006.1479;%%
\bibitem [{\citenamefont {Ma}(2007)}]{Ma:2007ku}%
  \BibitemOpen
  \bibfield  {author} {\bibinfo {author} {\bibfnamefont {E.}~\bibnamefont
  {Ma}},\ }\href {\doibase 10.1209/0295-5075/79/61001} {\bibfield  {journal}
  {\bibinfo  {journal} {Europhys.Lett.}\ }\textbf {\bibinfo {volume} {79}},\
  \bibinfo {pages} {61001} (\bibinfo {year} {2007})},\ \Eprint
  {http://arxiv.org/abs/hep-ph/0701016} {arXiv:hep-ph/0701016 [hep-ph]}
  \BibitemShut {NoStop}%
%%CITATION = HEP-PH/0701016;%%
\bibitem [{\citenamefont {Hagedorn}\ \emph {et~al.}(2009)\citenamefont
  {Hagedorn}, \citenamefont {Schmidt},\ and\ \citenamefont
  {Smirnov}}]{Hagedorn:2008bc}%
  \BibitemOpen
  \bibfield  {author} {\bibinfo {author} {\bibfnamefont {C.}~\bibnamefont
  {Hagedorn}}, \bibinfo {author} {\bibfnamefont {M.~A.}\ \bibnamefont
  {Schmidt}}, \ and\ \bibinfo {author} {\bibfnamefont {A.~Y.}\ \bibnamefont
  {Smirnov}},\ }\href {\doibase 10.1103/PhysRevD.79.036002} {\bibfield
  {journal} {\bibinfo  {journal} {Phys.Rev.}\ }\textbf {\bibinfo {volume}
  {D79}},\ \bibinfo {pages} {036002} (\bibinfo {year} {2009})},\ \Eprint
  {http://arxiv.org/abs/0811.2955} {arXiv:0811.2955 [hep-ph]} \BibitemShut
  {NoStop}%
%%CITATION = ARXIV:0811.2955;%%
\bibitem [{\citenamefont {BenTov}\ and\ \citenamefont
  {Zee}(2013)}]{BenTov:2012xp}%
  \BibitemOpen
  \bibfield  {author} {\bibinfo {author} {\bibfnamefont {Y.}~\bibnamefont
  {BenTov}}\ and\ \bibinfo {author} {\bibfnamefont {A.}~\bibnamefont {Zee}},\
  }\href {\doibase 10.1016/j.nuclphysb.2013.03.002} {\bibfield  {journal}
  {\bibinfo  {journal} {Nucl.Phys.}\ }\textbf {\bibinfo {volume} {B871}},\
  \bibinfo {pages} {452} (\bibinfo {year} {2013})},\ \Eprint
  {http://arxiv.org/abs/1202.4234} {arXiv:1202.4234 [hep-ph]} \BibitemShut
  {NoStop}%
%%CITATION = ARXIV:1202.4234;%%
\bibitem [{\citenamefont {Grimus}\ and\ \citenamefont
  {Ludl}(2010)}]{Grimus:2010ak}%
  \BibitemOpen
  \bibfield  {author} {\bibinfo {author} {\bibfnamefont {W.}~\bibnamefont
  {Grimus}}\ and\ \bibinfo {author} {\bibfnamefont {P.}~\bibnamefont {Ludl}},\
  }\href {\doibase 10.1088/1751-8113/43/44/445209} {\bibfield  {journal}
  {\bibinfo  {journal} {J.Phys.}\ }\textbf {\bibinfo {volume} {A43}},\ \bibinfo
  {pages} {445209} (\bibinfo {year} {2010})},\ \Eprint
  {http://arxiv.org/abs/1006.0098} {arXiv:1006.0098 [hep-ph]} \BibitemShut
  {NoStop}%
%%CITATION = ARXIV:1006.0098;%%
\bibitem [{\citenamefont {Mohapatra}\ and\ \citenamefont
  {Rodejohann}(2007)}]{Mohapatra:2006xy}%
  \BibitemOpen
  \bibfield  {author} {\bibinfo {author} {\bibfnamefont {R.}~\bibnamefont
  {Mohapatra}}\ and\ \bibinfo {author} {\bibfnamefont {W.}~\bibnamefont
  {Rodejohann}},\ }\href {\doibase 10.1016/j.physletb.2006.11.024} {\bibfield
  {journal} {\bibinfo  {journal} {Phys.Lett.}\ }\textbf {\bibinfo {volume}
  {B644}},\ \bibinfo {pages} {59} (\bibinfo {year} {2007})},\ \Eprint
  {http://arxiv.org/abs/hep-ph/0608111} {arXiv:hep-ph/0608111 [hep-ph]}
  \BibitemShut {NoStop}%
%%CITATION = HEP-PH/0608111;%%
\bibitem [{\citenamefont {Blum}\ \emph {et~al.}(2007)\citenamefont {Blum},
  \citenamefont {Mohapatra},\ and\ \citenamefont {Rodejohann}}]{Blum:2007qm}%
  \BibitemOpen
  \bibfield  {author} {\bibinfo {author} {\bibfnamefont {A.}~\bibnamefont
  {Blum}}, \bibinfo {author} {\bibfnamefont {R.}~\bibnamefont {Mohapatra}}, \
  and\ \bibinfo {author} {\bibfnamefont {W.}~\bibnamefont {Rodejohann}},\
  }\href {\doibase 10.1103/PhysRevD.76.053003} {\bibfield  {journal} {\bibinfo
  {journal} {Phys.Rev.}\ }\textbf {\bibinfo {volume} {D76}},\ \bibinfo {pages}
  {053003} (\bibinfo {year} {2007})},\ \Eprint {http://arxiv.org/abs/0706.3801}
  {arXiv:0706.3801 [hep-ph]} \BibitemShut {NoStop}%
%%CITATION = ARXIV:0706.3801;%%
\bibitem [{\citenamefont {Joshipura}\ and\ \citenamefont
  {Rodejohann}(2009)}]{Joshipura:2009fu}%
  \BibitemOpen
  \bibfield  {author} {\bibinfo {author} {\bibfnamefont {A.~S.}\ \bibnamefont
  {Joshipura}}\ and\ \bibinfo {author} {\bibfnamefont {W.}~\bibnamefont
  {Rodejohann}},\ }\href {\doibase 10.1016/j.physletb.2009.06.035} {\bibfield
  {journal} {\bibinfo  {journal} {Phys.Lett.}\ }\textbf {\bibinfo {volume}
  {B678}},\ \bibinfo {pages} {276} (\bibinfo {year} {2009})},\ \Eprint
  {http://arxiv.org/abs/0905.2126} {arXiv:0905.2126 [hep-ph]} \BibitemShut
  {NoStop}%
%%CITATION = ARXIV:0905.2126;%%
\bibitem [{\citenamefont {Gupta}\ \emph {et~al.}(2013)\citenamefont {Gupta},
  \citenamefont {Joshipura},\ and\ \citenamefont {Patel}}]{Gupta:2013it}%
  \BibitemOpen
  \bibfield  {author} {\bibinfo {author} {\bibfnamefont {S.}~\bibnamefont
  {Gupta}}, \bibinfo {author} {\bibfnamefont {A.~S.}\ \bibnamefont
  {Joshipura}}, \ and\ \bibinfo {author} {\bibfnamefont {K.~M.}\ \bibnamefont
  {Patel}},\ }\href@noop {} {\  (\bibinfo {year} {2013})},\ \Eprint
  {http://arxiv.org/abs/1301.7130} {arXiv:1301.7130 [hep-ph]} \BibitemShut
  {NoStop}%
%%CITATION = ARXIV:1301.7130;%%
\bibitem [{\citenamefont {Altarelli}\ \emph {et~al.}(2009)\citenamefont
  {Altarelli}, \citenamefont {Feruglio},\ and\ \citenamefont
  {Merlo}}]{Altarelli:2009gn}%
  \BibitemOpen
  \bibfield  {author} {\bibinfo {author} {\bibfnamefont {G.}~\bibnamefont
  {Altarelli}}, \bibinfo {author} {\bibfnamefont {F.}~\bibnamefont {Feruglio}},
  \ and\ \bibinfo {author} {\bibfnamefont {L.}~\bibnamefont {Merlo}},\ }\href
  {\doibase 10.1088/1126-6708/2009/05/020} {\bibfield  {journal} {\bibinfo
  {journal} {JHEP}\ }\textbf {\bibinfo {volume} {0905}},\ \bibinfo {pages}
  {020} (\bibinfo {year} {2009})},\ \Eprint {http://arxiv.org/abs/0903.1940}
  {arXiv:0903.1940 [hep-ph]} \BibitemShut {NoStop}%
%%CITATION = ARXIV:0903.1940;%%
\bibitem [{\citenamefont {de~Adelhart~Toorop}\ \emph
  {et~al.}(2010)\citenamefont {de~Adelhart~Toorop}, \citenamefont {Bazzocchi},\
  and\ \citenamefont {Merlo}}]{Toorop:2010yh}%
  \BibitemOpen
  \bibfield  {author} {\bibinfo {author} {\bibfnamefont {R.}~\bibnamefont
  {de~Adelhart~Toorop}}, \bibinfo {author} {\bibfnamefont {F.}~\bibnamefont
  {Bazzocchi}}, \ and\ \bibinfo {author} {\bibfnamefont {L.}~\bibnamefont
  {Merlo}},\ }\href {\doibase 10.1007/JHEP08(2010)001} {\bibfield  {journal}
  {\bibinfo  {journal} {JHEP}\ }\textbf {\bibinfo {volume} {1008}},\ \bibinfo
  {pages} {001} (\bibinfo {year} {2010})},\ \Eprint
  {http://arxiv.org/abs/1003.4502} {arXiv:1003.4502 [hep-ph]} \BibitemShut
  {NoStop}%
%%CITATION = ARXIV:1003.4502;%%
\bibitem [{\citenamefont {Patel}(2011)}]{Patel:2010hr}%
  \BibitemOpen
  \bibfield  {author} {\bibinfo {author} {\bibfnamefont {K.~M.}\ \bibnamefont
  {Patel}},\ }\href {\doibase 10.1016/j.physletb.2010.11.024} {\bibfield
  {journal} {\bibinfo  {journal} {Phys.Lett.}\ }\textbf {\bibinfo {volume}
  {B695}},\ \bibinfo {pages} {225} (\bibinfo {year} {2011})},\ \Eprint
  {http://arxiv.org/abs/1008.5061} {arXiv:1008.5061 [hep-ph]} \BibitemShut
  {NoStop}%
%%CITATION = ARXIV:1008.5061;%%
\bibitem [{\citenamefont {Meloni}(2011)}]{Meloni:2011fx}%
  \BibitemOpen
  \bibfield  {author} {\bibinfo {author} {\bibfnamefont {D.}~\bibnamefont
  {Meloni}},\ }\href {\doibase 10.1007/JHEP10(2011)010} {\bibfield  {journal}
  {\bibinfo  {journal} {JHEP}\ }\textbf {\bibinfo {volume} {1110}},\ \bibinfo
  {pages} {010} (\bibinfo {year} {2011})},\ \Eprint
  {http://arxiv.org/abs/1107.0221} {arXiv:1107.0221 [hep-ph]} \BibitemShut
  {NoStop}%
%%CITATION = ARXIV:1107.0221;%%
\end{thebibliography}%

\end{document}